\documentclass{article}   	% use "amsart" instead of "article" for AMSLaTeX format
\usepackage[english]{babel}
\usepackage{geometry}                		% See geometry.pdf to learn the layout options. There are lots.
\usepackage[parfill]{parskip}    		% Activate to begin paragraphs with an empty line rather than an indent
\usepackage{amssymb,bm,amsmath,amsfonts}
\usepackage{multicol}
\usepackage{lipsum,float}
\usepackage{graphicx}
\usepackage{authblk}
\usepackage{soul}
\usepackage{subcaption}
\title{\textbf{Saturation of fishbone instability by self-generated zonal flows in tokamak plasmas}}
\author[1,2]{G. Brochard\thanks{Corresponding authors: \texttt{guillaume.brochard@iter.org, zhihongl@uci.edu}}}
\author[3]{C. Liu}
\author[2]{X. Wei}
\author[2]{W. Heidbrink}
\author[*2]{Z. Lin}
\author[3]{N. Gorelenkov} 
\author[4]{C. Chrystal}
\author[4]{X. Du}
\author[5]{J. Bao}
\author[1]{A. R. Polevoi}
\author[1]{M. Schneider}
\author[1]{S. H. Kim}
\author[1]{S. D. Pinches}
\author[2]{P. Liu}
\author[2]{J. H. Nicolau}
\author[6]{H. L\"utjens}
\affil[1]{\small{\emph{ITER organisation, Route de Vinon-sur-Verdon, CS 90 046 13067 St., Paul Lez Durance, France}}}
\affil[2]{\small{\emph{Department of Physics and Astronomy, University of California, Irvine, California 92697, USA}}}
\affil[3]{\small{\emph{Princeton Plasma Physics Laboratory, Princeton University, PO Box 451, Princeton, New Jersey 08543,USA}}}
\affil[4]{\small{\emph{General Atomics, PO Box 85608, San Diego, CA 92186-5608, United States of America}}}
\affil[5]{\small{\emph{Institute of Physics, Chinese Academy of Sciences, Beijing 100190, China}}}
\affil[6]{\small{\emph{CPHT, CNRS, \'Ecole Polytechnique, Institut Polytechnique de Paris, Route de Saclay, 91128 Palaiseau, France}}}
\date{}							% Activate to display a given date or no date

\begin{document}
\thispagestyle{empty}
\maketitle
%\section{}
%\subsection{}
\begin{abstract}
\noindent
Gyrokinetic simulations of the fishbone instability in DIII-D tokamak plasmas find that self-generated zonal flows can dominate the nonlinear saturation by preventing coherent structures from persisting or drifting in the energetic particle phase space when the mode frequency down-chirps. Results from the simulation with zonal flows agree quantitatively, for the first time, with experimental measurements of the fishbone saturation amplitude and energetic particle transport. Moreover, the fishbone-induced zonal flows are likely responsible for the formation of an internal transport barrier that was observed after fishbone bursts in this DIII-D experiment. Finally, gyrokinetic simulations of a related ITER baseline scenario show that the fishbone induces insignificant energetic particle redistribution and may enable high performance scenarios in ITER burning plasma experiments.

\end{abstract}
\begin{figure}[H]
\begin{center}
   \includegraphics[scale=0.25]{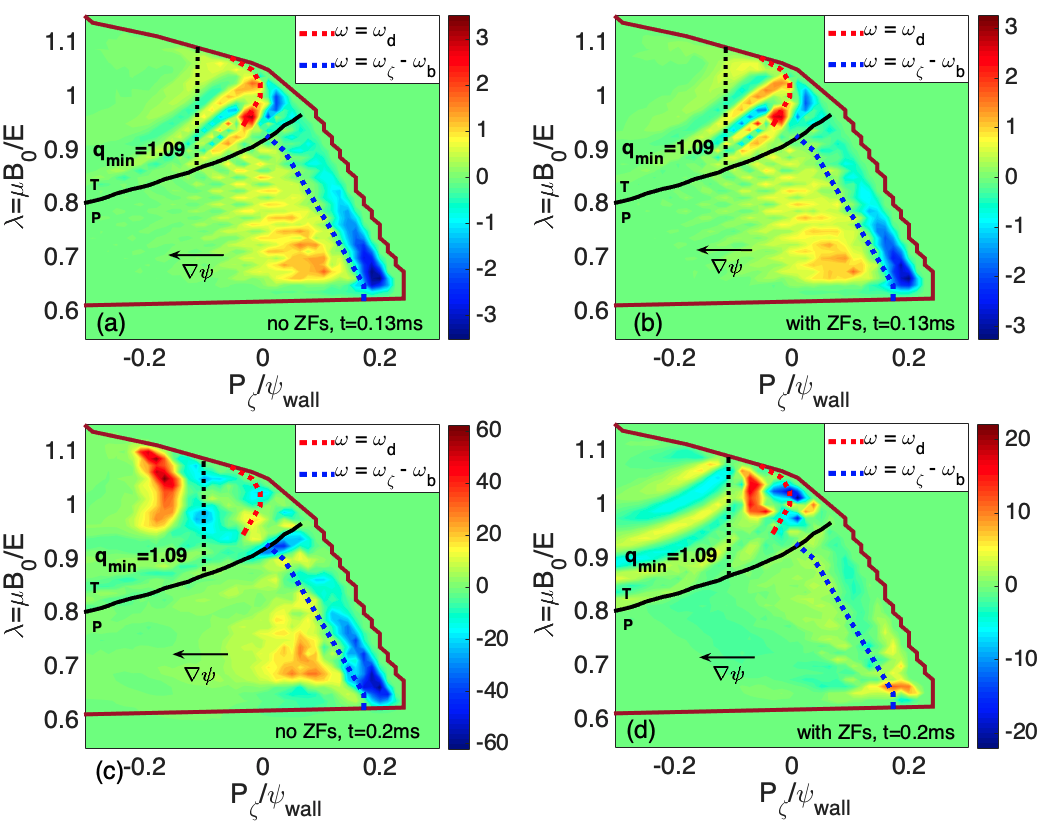}
\end{center}
\caption{Instantaneous EP distribution $\partial_t\delta f$ in linear (top) and nonlinear (bottom) phases, without (left) and with (right) zonal flows in GTC simulations. Passing (P) and trapped (T) phase space zones are separated by a black line, and envelope of EP distribution at $\mu B_0 = 45$ keV is highlighted by a brown line.}
\label{res}
\end{figure}

\begin{multicols}{2}
\emph{Introduction.} - Energetic Particles (EPs) in fusion plasmas can destabilize instabilities at different spatial scales that may lead to their outward transport. This is a critical issue for burning plasmas as in ITER \cite{ITER}, where the core confinement of 3.5 MeV fusion-born alphas and other EPs is essential to maintain a dominantly self-heated plasma with a fusion gain of Q$\geq$10. This EP transport therefore needs to be predicted for mitigation strategies to be incorporated in plasma scenarios. \\
The macroscopic MHD mode named fishbone \cite{McGuire1983}\cite{Chen1984}, a helicoidal EP-driven instability of the core plasma in tokamaks, is one such instability that could drive large EP transport in fusion devices. Its amplitude and the associated EP transport depend strongly on nonlinear processes that lead to the mode saturation. The mechanism presumably dominating the fishbone saturation was identified in kinetic-MHD simulations \cite{Fu2006}\cite{Brochard2020b} to be the resonant wave-particle interaction. However self-generated zonal flows (ZFs), which are axisymmetric flows due to the electrostatic potential averaged over the magnetic flux surfaces, were not included or under-evaluated in these simulations without kinetic thermal ions, which can impact the fishbone saturation \cite{Chen2016}. Indeed, zonal flows are known to dominate the saturation of instabilities arising at microscopic and mesoscopic scales such as drift-waves \cite{Lin1998} and Alfv\'en eigenmodes (AEs) \cite{Spong1994}\cite{Chen2012}. The radial electric field shear of zonal flows can suppress turbulent transport \cite{Hahm1995}, resulting in the formation of an internal transport barrier (ITB) that greatly enhances plasma confinement \cite{Diamond2005}. An outstanding issue is whether zonal flows can play a similar role in saturating macroscopic MHD modes, which may affect turbulent transport through cross-scale interactions common in fusion \cite{Liu2022a} and astrophysical plasmas \cite{Giacalone1999}.
 \\
In this Letter, we report the first self-consistent gyrokinetic simulations finding fishbone saturation by self-generated zonal flows, in a DIII-D tokamak experiment \cite{Heidbrink2021} which is chosen for experimental comparison purposes to predict the EP transport in an ITER pre-fusion power operation (PFPO) scenario \cite{Polevoi2021}. Global gyrokinetic simulations using the GTC code \cite{Lin1998} find that self-generated zonal flows greatly reduce the fishbone saturation amplitude, by preventing the EP phase space zonal structures \cite{Zonca2015}\cite{Falessi2019} from persisting or drifting in the nonlinear phase with the mode frequency down-chirping, which reduces the EP resonant drive. Results from the simulation with zonal flows agree quantitatively, for the first time, with experimental measurements of the fishbone saturation amplitude and the neutron emissivity drop due to the associated EP transport. Moreover, the shearing rate of the zonal electric field exceeds the linear growth rate of unstable drift-waves. The potential suppression of microturbulence by fishbone-induced zonal flows is likely the mechanism for the formation of an internal transport barrier (ITB) that was observed after fishbone bursts in this DIII-D experiment, similar to that in the EAST tokamak \cite{Ge2022}. Our simulation results confirm the long suspected role of fishbones in ITB formations \cite{Pinches2001}, where fishbone bursts have been observed to precede ITB formations in ASDEX \cite{Guenter2001}, MAST \cite{Field2011}, HL-2A \cite{Chen2016a} and EAST \cite{Gao2018} plasmas. Finally, after having validated GTC for fishbone modes with this DIII-D experiment, gyrokinetic simulations find that the fishbone-induced EP transport in the related ITER baseline scenario is insignificant, with less than 2\% of the core EPs being redistributed. The intentional destabilization of fishbone modes in ITER is therefore possibly a way to enhance fusion performances. \\
\emph{Experimental setup.} - The selected DIII-D discharge \#178631 \cite{Heidbrink2021} has an oval shape (elongation $\kappa=1.17$, triangularity $\delta=0.07$) that is limited on the carbon inner wall. The
major radius is $R_0=1.74$ m, the minor radius is $a = 0.64$ m, the toroidal field is
$2.0$ T, the plasma current is $0.88$ MA, and the line-average electron density is $\sim2.0\times10^{19}$ m$^{-3}$. This discharge was chosen primarily because it has an accurately known, weakly reversed safety factor profile with $q_0=1.2$, $q_{min}=1.09$, and $q_{95}=3.8$ values that resembles the profile predicted for the ITER PFPO baseline scenario. The deuterium, L-mode plasma is heated by 3.8 MW of 81 keV deuterium beams that are injected in the midplane in the direction of the plasma current and by 1.0 MW of 2$^{nd}$ harmonic, central electron cyclotron heating.
 \\
\emph{Numerical setups.} - The DIII-D discharge \#178631 is studied numerically through gyrokinetic simulations with the GTC code \cite{Lin1998}, and with kinetic-MHD simulations using the M3D-C1 \cite{Liu2022} and XTOR-K \cite{Brochard2020a} codes. The magnetic configuration at t=1580ms is reproduced from the EFIT code \cite{Lao1985}. Plasma profiles are obtained from TRANSP \cite{Grierson2018} simulations using experimental measurements. To ensure that the sum of pressures of all species from TRANSP equals to the total pressure in EFIT, the EP pressure is constrained as $p_{f}=p_{tot}-p_{i}-p_{e}$, given that the uncertainty on EP profiles is the highest in TRANSP. The experimental NBI distribution  is reproduced from the NUBEAM code \cite{Pankin2004}. It is then fitted analytically with a slowing-down model \cite{Moseev2019} taking into account the three injections energies resulting from the NBI positive-ion sources in DIII-D. All nonlinear simulations cover the core region including the magnetic axis, with an outer edge buffer after $\rho_T=\sqrt{\psi_T/\psi_{T,edge}} =0.8$ in GTC where equilibrium gradients are removed, with $\psi_T$ the poloidal flux. GTC gyrokinetic simulations retain only the n=1 mode, with or without the n=m=0 zonal flows, and use gyrokinetic thermal/fast ions with a $\delta f$ method and massless fluid electrons \cite{Lin2001}. Electron contribution to zonal density is neglected based on their adiabatic response, in order to avoid numerical instability. Nonetheless, its effects need to be studied in a future study. M3D-C1 kinetic-MHD simulations incorporate low-n modes $n=0,1$ with both thermal and fast ions kinetic effects \cite{Liu2022b} using a $\delta f$ method. XTOR-K kinetic-MHD simulation treats kinetically only the fast ion species in this work with a full-f method. Due to the anisotropic nature of the chosen configuration that has $\beta_f/\beta_{tot}=54\%$ on axis, where $\beta$ is the ratio between thermal and magnetic pressures, there is  a significant evolution of the n=0 kinetic-MHD equilibrium in XTOR-K, which departs substantially from the isotropic initial EFIT reconstruction used in the $\delta f$ codes. For this reason, the $n=0$ mode is filtered out in the present study with XTOR-K, only considering the $n=1$ mode. Convergence studies over spatial grid size, time step, number of particles per cell and radial boundary treatment have been successfully conducted, with $N_{\psi}=100$, $N_{\theta}=250$, $N_{\parallel}=24$ and a Gaussian boundary decay from $\rho_T=0.73$ to $\rho_T=1$ in GTC fishbone simulations. These three codes have been verified and  linearly validated for simulations of the n=1 kink instability in another DIII-D experiment \cite{Brochard2022}, demonstrating the capability of GTC's gyrokinetic formulation at simulating n=1 MHD modes.  Fishbone modes in these DIII-D/ITER plasmas respect the usual gyrokinetic ordering $k_{\parallel}/k_{\perp}\ll 1$, as for these modes $k_{\parallel}\sim 0$ and $k_{\perp}^{-1}\sim r_{q_{min}}\sim 0.3a$.\\
\emph{Fishbone saturation by self-generated zonal flows -} Both gyrokinetic and kinetic-MHD simulations find that the internal kink mode is stable in the absence of EPs, and that the fishbone is driven unstable by EPs in this DIII-D experiment, with a marginal stability threshold at $p_{f,thres} = 0.8 p_{f}$. The unstable fishbone has a growth rate of $\gamma_{n=1}=8.5\times10^4$ s$^{-1}$ and a mode frequency of $\omega/2\pi = 17$kHz in GTC simulations. When the realistic beam distribution is replaced by its equivalent Maxwellian distribution, this mode is fully stabilized, showing the sensitivity of fishbone instabilities to EP distributions. The effects of MHD nonlinearities on the n=1 fishbone were previously examined in kinetic-MHD simulations by keeping side-band n=0-4 modes, indicating reduction of initial saturation amplitude \cite{Fu2006}\cite{Ge2022} by additional dissipation, and generation of n=m=0 zonal flows \cite{Brochard2020b}\cite{Ge2022}. The specific role played by zonal flows in the fishbone saturation was however not identified, and kinetic thermal ions were not included. 
 \begin{figure}[H] 
 \begin{center}
   \includegraphics[scale=0.23]{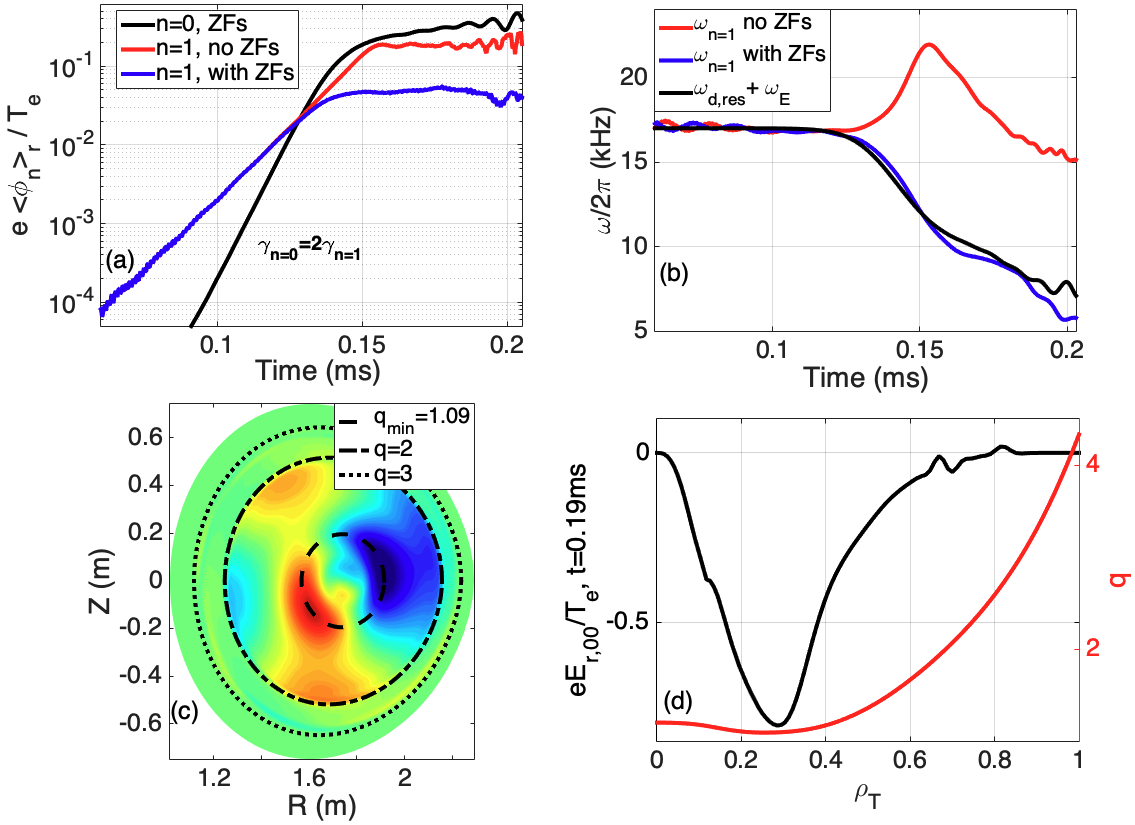}
   \end{center}
\caption{ Time evolution of (a) volume-averaged perturbed electrostatic potential $e\phi/T_e$ (n=0,1), (b) n=1 mode frequency $\omega_{n=1}$ and linearly resonant precessional frequency $\omega_{d,res}$ plus zonal $E\times B$ frequency $\omega_E$ at $q_{min}$, in GTC simulations. (c) $n=1 \ e\phi/T_e$ in poloidal plane and (d) Radial profiles of safety factor q and zonal electric field $eE_{r,00}/T_e$ (a.u.) after saturation at t=0.19ms.}
\label{hist}
\end{figure}
The effects of zonal flows on the fishbone instability are studied here self-consistently for the first time with gyrokinetic simulations. A gyrokinetic treatment of thermal ions is essential for a realistic description of zonal flows in both kinetic-MHD and gyrokinetic codes, to account for their residual level after collisionless damping \cite{Rosenbluth1998}. GTC nonlinear simulations are performed with and without the n=m=0 zonal flows, as illustrated in Fig.\ref{hist}. The time evolution of the volume-averaged electrostatic potential $e\phi/T_e$ shows that n=m=0 zonal flows are generated during the linear excitation of the fishbone mode, with exactly $\gamma_{n=0}=2\gamma_{n=1}$. Given that only $|n|=0,1$ modes are retained in the simulation, zonal flows are necessarily driven by the $n=\pm1$ fishbone throught mode-mode coupling. Such dynamics is similar to what was observed numerically \cite{Todo2012} and found analytically for TAEs (Toroidal Alfv\'en Eigenmodes), where zonal flows generation is force-driven \cite{Qiu2016} rather than through modulational instability \cite{Chen2000}. The n=1 fishbone saturates near $t\sim0.15$ms with $\delta B/B_0\sim2\times10^{-3}$ with zonal flows, and at $\delta B/B_0\sim8\times10^{-3}$ without. Zonal flows also reduce significantly the EP diffusivity at saturation, from 30 to 4 m$^2$/s. After fishbone saturation, zonal flows amplitude continues to increase slowly, which has also been commonly observed in gyrokinetic simulations of microturbulence and Alfv\'en eigenmode \cite{Liu2022a}.
\begin{figure}[H]
\begin{center}
\includegraphics[scale=0.245]{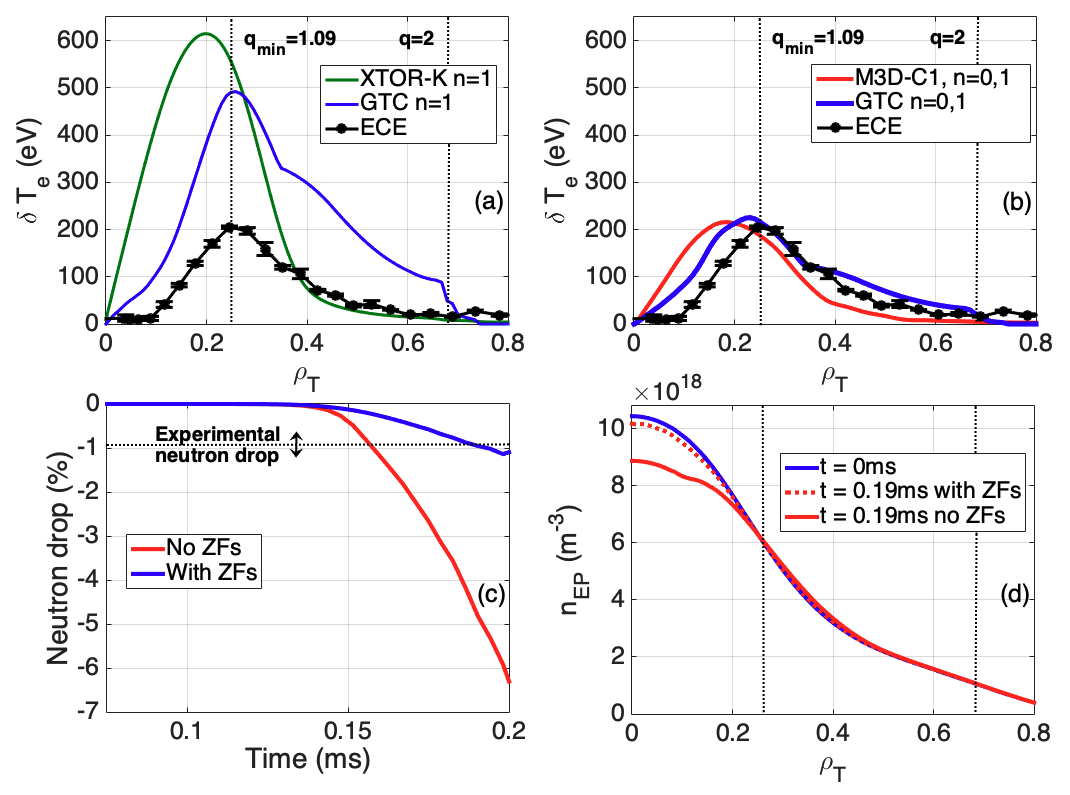}
\end{center}
\caption{Radial envelope of electron temperature perturbation $\delta T_e$ after saturation without (a) and with (b) zonal flows in GTC, M3D-C1 and XTOR-K simulations, compared to ECE measurement in DIII-D \#178631. (c) Time evolution of neutron drop $\delta\Gamma_N$ in GTC. (d) EP density profiles in GTC simulations before and after fishbone burst. }
\label{val}
\end{figure} 
Microturbulence and meso-scale MHD modes, discarded in this work, can impact the fishbone nonlinear dynamics by driving or damping zonal flows and scattering EPs, which will be studied in future cross-scale simulations. Collisions, which can also affect the zonal flows levels, are neglected in GTC simulations based on the short simulation time ($\Delta t\sim2\times10^{-4}$s) compared to the ion-ion collision time ($\tau_{ii}\sim3\times10^{-2}$s ).\\
As shown in Figure \ref{hist}b, the mode frequency down-chirps in the laboratory frame after the n=1 mode saturation, which is a typical fishbone signature. In the simulation without zonal flows, the mode frequency chirps up right before the fishbone saturation but chirps down when the mode starts saturating. This initial up-chirping may be attributed to the larger mode amplitude near saturation that induces ideal MHD nonlinear effects \cite{Cowley1996}. The down-chirping occurs independently of the initial perturbation chosen in the simulations. The n=1 electrostatic potential and the n=0 radial electric field after saturation at t=0.19ms are displayed on Fig.\ref{hist}c-d. The n=1 mode features a dominant m=1 harmonic centered around $q_{min}$, as well as a significant m=2 side-band that vanishes outside the $q=2$ surface. The zonal electric field exhibits a macroscopic structure centered near $q_{min}$ as well, which differs from the usual microscospic/mesoscopic scales associated with zonal flows generated by drift-wave modes/AEs \cite{Diamond2005}\cite{Chen2016}. This electric field leads to a strongly sheared poloidal rotation in the electron diamagnetic direction, which is opposite to the n=1 fishbone propagation direction. \\ 
This novel mechanism of fishbone saturation by self-generated zonal flows is supported by DIII-D measurements of the fishbone amplitude shown in Fig.\ref{val}. The electron temperature perturbation envelope $\delta T_e$ obtained from GTC, M3D-C1 and XTOR-K nonlinear simulations at saturation are compared with the ECE measurements \cite{Austin2003} on  Fig.\ref{val} (a-b). The $\delta T_e$ envelope is defined here as the sum of all poloidal harmonics for the $n=1$ mode.  The ECE measurements represent here the maximal amplitude of the fishbone mode during its burst. Since in our simulations the $n=1$ mode is fully saturated, we expect that the simulated amplitudes represent the experimentally measured maximal fishbone amplitude during the early phase of a complete fishbone burst cycle, allowing comparisons on shorter time scales. Without zonal flows, XTOR-K and GTC results have comparable saturation amplitudes with $\delta T_{e,max}\sim500-600$ eV, which are three time larger than the experimental value. When including zonal flows, GTC saturation amplitude at $\delta T_{e,max}\sim200$ eV matches well with the experimental one. M3D-C1 saturation amplitude also agree reasonably with both GTC and the ECE measurements, qualitatively supporting the nonlinear gyrokinetic simulation of the fishbone mode. The significant m=2 harmonic outside the $q_{min}$ surface in the GTC simulation leads to a quantitative agreement with the ECE measurement, which provides a successful comparison between GTC simulations and DIII-D experimental measurements. Nonlinear simulation scans over the radial position and amplitude of $q_{min}$ using GTC all show the same significant saturation by zonal flows. This good agreement is further demonstrated by comparing the simulated and experimental neutron emissivities. In GTC the perturbed volume-averaged neutron flux is defined as $\delta\Gamma_{N} = \langle n_{i}\delta f_{EP}\sigma v\rangle_r = \langle n_i\sum_k\delta(\textbf{x}-\textbf{x}_k)\delta(\textbf{v}-\textbf{v}_k)\sigma(v_k)v_k\rangle_r$ with $n_i$ the thermal ion density profile, $\delta f_{EP}$ the EP perturbed distribution, $\textbf{x}_k$ and $\textbf{v}_k$ the position and velocity of each EP markers, and $\sigma$ the D-D nuclear fusion cross section. As shown on Fig.\ref{val}c, without zonal flows GTC exhibits a neutron drop of $\delta\Gamma_N\sim$ 6\%, much higher than the experimental one at $\delta\Gamma_N=0.9\%\pm0.3\%$. When including zonal flows in the simulation however, the neutron drop yields $\delta \Gamma_N\sim1.1\%$, which agrees with the measurements within the experimental uncertainty. The simulated $\delta\Gamma_n$ levels have however not reached a steady-state since our simulations do not complete a full fishbone cycle due to numerical constraints. It could imply that our neutron drop levels are slightly under-estimated. As expected from these neutron drop values, the fishbone-induced EP transport from the simulation with zonal flows is rather weak as shown on Fig. \ref{val}d, with about 3\% of EPs inside of the $q_{min}$ surface redistributed outward. The redistribution is more significant in the simulation without zonal flows, as it affects 15\% of EPs in the core plasma. \\
\emph{Mechanism for fishbone saturation by zonal flows} - The EP phase-space analysis reveals that zonal flows influence the time evolution of coherent phase space zonal structures, thus impacting the n=1 fishbone mode saturation. On Fig.\ref{res}, the instantaneous EP distribution evolution $\partial_t\delta f$ is displayed in the phase space diagram $(P_{\zeta},\lambda=\mu B_0/E)$, with $P_{\zeta}$ and $\lambda$ respectively the toroidal canonical momentum and the pitch angle, at fixed magnetic moment $\mu B_0=45$keV before and after the fishbone saturation. The time derivative is chosen rather than the usual perturbed EP distribution $\delta f$ \cite{Brochard2020b} since the fishbone mode frequency is chirping in the nonlinear phase, which causes phase space zonal structures to drift in time.
In the linear phase, the mode is excited by two resonances, the precessional one $\omega=\omega_d$ due to trapped particles, and a drift-transit one due to passing particles $\omega=\omega_{\zeta}-\omega_{b}$ with $\omega_{\zeta}$, $\omega_{b}$ the toroidal and poloidal transit frequencies, similar to \cite{Brochard2020b}. As can be observed on Fig.\ref{res} (a-b), a hole and clump structure develops around each resonances in the late linear phase in both cases, indicating a resonant outward EP redistribution.
In the nonlinear phase shown in Fig.\ref{res}(c-d), the dynamical evolution of these phase space zonal structures differ significantly between the simulations with and without zonals flows. Without zonal flows, the hole and clump in the trapped region (around the red dotted line in Fig.\ref{res}) nonlinearly drift to higher $\psi$ positions due to the mode down-chirping since $\omega_d\propto1/\sqrt{\psi}$, while the one in the passing part does not move. However with zonal flows, the phase space zonal structure in the trapped region remains static, even though the mode is chirping down. The hole and clump in the passing particle region vanishes. Therefore, zonal flows nonlinearly prevent the resonant interaction of the fishbone with EPs that are not linearly resonant with the mode, which leads to a weaker fishbone saturation due to the absence of additional resonant drive. This reduction is illustrated on Fig.\ref{res}(c-d) by the lower amplitude of the hole and clump in the trapped region with zonal flows. This novel physics of phase space zonal structures trapping by fishbone-induced zonal flows is reminiscent of the trapping of turbulence eddies by zonal flows in microturbulence \cite{Diamond2005}\cite{Guo2009}.  \\ These differences in nonlinear evolution can be explained by the influence of zonal flows on the EP wave-particle resonance. The perturbed radial electric field associated with zonal flows generates a $E\times B$ frequency $\omega_{E}=\delta\textbf{v}_{E}\cdot(mq\nabla\theta-n\nabla\zeta)$, with $\delta\textbf{v}_E=\textbf{b}_0\times\nabla\phi_{00}/B_0$, which modifies the precessional resonance condition to $\omega=\omega_d+\omega_E$, as discussed in \cite{Brochard2020b} and \cite{Chen2016} (Eq. 4.182). As can be observed on Fig.\ref{hist}b, the time evolution of the precessional frequency of linearly resonant EPs plus the perturbed $E\times B$ frequency at $\rho=\rho_{q_{min}}$ matches almost exactly the time evolution of the fishbone frequency with zonal flows. Therefore the linear resonance condition is still satisfied despite the mode down-chirping in laboratory frame, which explains why the phase space zonal structure in the trapped region remains static. The $E\times B$ flow shear could also modify the EPs transit frequencies, which may account for the disappearance of the $\omega=\omega_{\zeta}-\omega_b$ hole and clump through resonance detuning. Zonal flows are therefore able to dominate the fishbone saturation by strongly reducing the resonant wave-particle drive. \\
\begin{figure}[H]
\begin{center} 
   \includegraphics[scale=0.18]{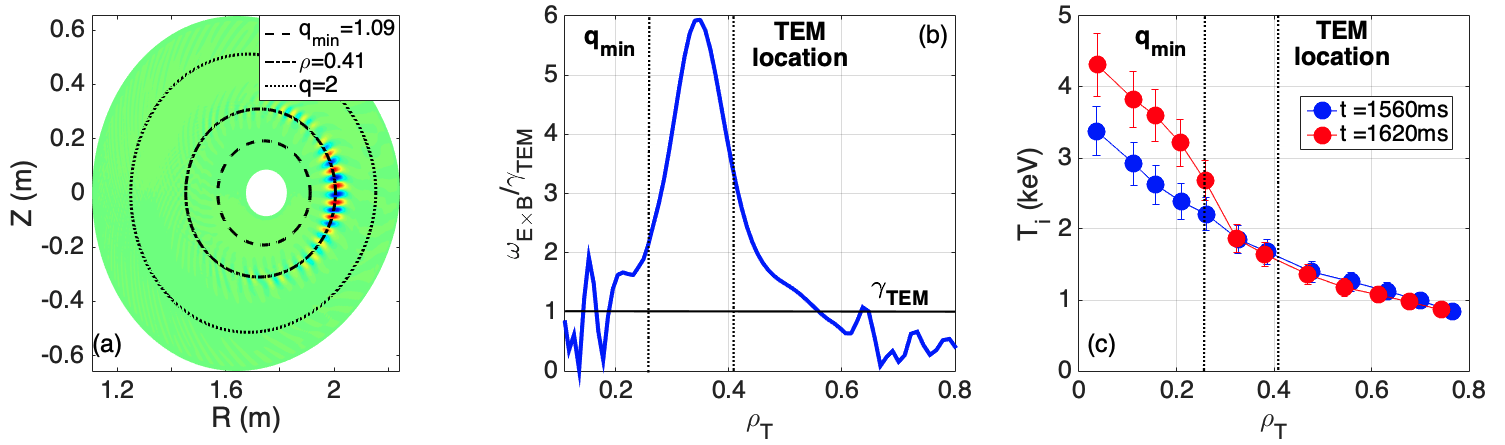}
\end{center}
\caption{a) Electrostatic potential $\phi$ of unstable TEM in the poloidal plane in GTC simulation. b) Fishbone-induced shearing rate profile after saturation. c) $T_i$ profiles before and after fishbone bursts from CXRS measurements in DIII-D shot \#178631.}
\label{ITB}
\end{figure}
\emph{Fishbone-induced ITB formation} - Besides affecting the fishbone saturation, zonal flows also generate a strong shear within $\rho_T\in[0.1,0.5]$, with an instantaneous shearing rate of $\gamma_E\sim3\times10^5$ s$^{-1}$. High-n electrostatic GTC simulations with kinetic trapped electrons were performed for this DIII-D configuration, finding that the most unstable drift-wave is a collisionless trapped electron mode (CTEM) at $\rho=0.4$, shown on Fig.\ref{ITB}a, with a linear growth rate of $\gamma_{TEM}=1.38\times 10^5$ s$^{-1}$ and a wavelength of $k_{\theta}\rho_i \sim 0.5$. A resolution of $N_{\psi}=120, N_{\theta}=1200, N_{\parallel}=32$ was used in this simulation, with the toroidal modes domain $n\in[30,50]$, the TEM peaking at $n\sim40$. Since the shearing rate generated by the fishbone is larger than the TEM growth rate and that the ratio of TEM radial to poloidal wavelength is much larger than one, the effective shearing rate of the fishbone-induced zonal flows is much larger than the TEM growth rate. Therefore, the zonal flows generated by the fishbone could suppress the turbulence \cite{Hahm1995}, confirming the speculated role of fishbones in the formation of ITBs \cite{Pinches2001}. Evidence of microturbulence suppression is obtained experimentally in DIII-D with the ion temperature measurement using the charge exchange recombination spectroscopy (CXRS) diagnostic \cite{Chrystal2016}. The formation of an ion ITB after fishbone bursts occurring at t=1581,1594,1607 and 1615ms can indeed be observed in Fig.\ref{ITB} c. The increase of the core $T_i$ cannot be explained by the heating from the beam, as it was at constant power since t=300ms, for multiple slowing-down times before the onset of fishbones. Fishbone bursts were also observed to precede ITBs in four other DIII-D discharges with similar heating power, density, current and $q_{min}$ parameters.  \\
 \emph{EP transport in ITER PFPO scenario} - Building on the good agreement with DIII-D experimental measurements, GTC is now applied to the selected ITER PFPO scenario with a 7.5 MA/2.65 T H-mode plasma heated by 33 MW of NBI and 20 MW of ECH \cite{Polevoi2021}, to predict the fishbone-induced EP transport. Similar to the DIII-D simulations, the EP beam is fitted from an analytical anisotropic slowing-down distribution. Linear GTC simulations show that the n=1 fishbone is unstable, with a mode growth rate and frequency of $\gamma=4.4\times10^4$ s$^{-1}$ and $\omega/2\pi=48$ kHz, while simulations with equivalent Maxwellian distributions find a stable n=1 mode. \\
Similarly to DIII-D simulations, zonal flows lower the n=1 fishbone saturation amplitude. The zonal electric field at the fishbone saturation  peaks with negative values close to the $q_{min}$ surface. Electrostatic GTC simulations were also performed for this ITER scenario, finding an unstable TEM at $\rho=0.71$ with $\gamma_{TEM}=3\times10^4$ s${^{-1}}$ using $N_{\psi}=500, N_{\theta}=3600, N_{\parallel}=32$  $n\in[100,250]$, the TEM peaking at $n\sim170$. At that location, the fishbone-induced shearing rate is 200\% larger than the TEM linear growth rate, suggesting that an ITB could also be triggered for this ITER PFPO scenario. \\ 
After the end of the fishbone burst in the simulation without zonal flows, the overall redistribution within $q_{min}$ is of about 2\% of the initial distribution, which gives an upper-bound for the EP redistribution by the fishbone. Overall, the NBI fishbone should not impact significantly the plasma heating of this ITER PFPO scenario, similar to what was shown for the alpha-fishbone in the ITER 15MA baseline DT scenario \cite{Brochard2020b}. \\ Longer cross-scale GTC simulations will be necessary in future works to confirm microturbulence suppression by the fishbone instability in these DIII-D and ITER cases \cite{Liu2022a}.\\
\emph{Discussion} - Since fishbone oscillations may not cause significant EP redistribution in ITER plasmas, it can be of great interest to design ITER scenarios triggering them on purpose rather than avoiding them, by tailoring NBI and ICRH depositions to excite fishbone resonances. As was shown in this Letter, fishbones can generate strongly sheared flows that can damp drift-wave instabilities and hence reduce the turbulent transport. While it was observed here and in several other tokamaks \cite{Guenter2001}\cite{Field2011}\cite{Chen2016a}\cite{Gao2018} that fishbone oscillations can lead to ITB formations, that was not the case in some others such as JET \cite{Joffrin2002}, despite efforts to reproduce the fishbone-induced ITB formations observed in ASDEX plasmas \cite{Guenter2001}.  There may therefore exist parametric dependencies (q profile, EP pressure profile, trapped/passing EP ratio) for the fishbone instability that control the emergence of strongly sheared zonal flows \cite{Wolf2003}. The computational prediction and experimental observation of such dependencies could enable the creation of high-performance scenarios, of crucial importance for ITER burning plasmas.\\
The views and opinions expressed herein do not necessarily reflect those of the ITER organization. This work was supported by DOE SciDAC ISEP, INCITE, and the ITPA-EP group. This material is based upon work supported by the U.S. DOE, Office of Science, Office of Fusion Energy Sciences, using the DIII-D National Fusion Facility, a DOE Office of Science user facility, under Award(s) DE-FC02-04ER54698, as well as computing resources from ORNL (DOE Contract DE-AC05-00OR22725), NERSC (DOE Contract DE- AC02-05CH11231), and PPPL (DOE Contract DE-AC02-09CH11466).
\bibliography{fishbone_zonal_prl}{}
\end{multicols}
\end{document}